\documentclass[11pt]{article}
\usepackage[a4paper,hmarginratio=1:1,vmarginratio=2:3,
totalwidth=15.1cm,totalheight=22.2cm]{geometry}
\usepackage{bm,epstopdf,epsfig,amsmath,amssymb,amsfonts,colordvi,latexsym,comment,cancel,
verbatim}
\usepackage{graphicx,graphics}
\usepackage[font=md,captionskip=8pt]{subfig}
\usepackage[usenames,dvipsnames]{color}

\usepackage{setspace}

\newcommand{\cO}{{\cal O}}

\newcommand{\ra}{\rightarrow}
\newcommand{\be}{\begin{equation}}
\newcommand{\ee}{\end{equation}}
\newcommand{\bea}{\begin{eqnarray}}
\newcommand{\eea}{\end{eqnarray}}

\newcommand{\baa}{\begin{array}}
\newcommand{\eaa}{\end{array}}
\long\def\symbolfootnote[#1]#2{\begingroup
\def\thefootnote{\fnsymbol{footnote}}\footnote[#1]{#2}\endgroup}
\begin{document}
\begin{flushright}
CERN-PH-TH/2013-033.\\
\today
\end{flushright}

\thispagestyle{empty}

\vspace{1.5cm}

\begin{center}
{\Large {\bf Fixing the EW scale in supersymmetric  models 

\bigskip
after the Higgs discovery.}}\\
\medskip
\vspace{1.cm}
\textbf{
D.~M. Ghilencea$^{\,a,\,b,\,}$\symbolfootnote[1]{E-mail address: 
dumitru.ghilencea@cern.ch}} \\

\bigskip

$^{a\,}${\small CERN - Theory Division, CH-1211 Geneva 23, Switzerland.}

$^{b\,}${\small Theoretical Physics Department, National Institute of Physics }

{\small and Nuclear Engineering (IFIN-HH) Bucharest MG-6 077125, 
Romania.}

\end{center}

\def\baselinestretch{1.14}
\begin{abstract}
TeV-scale supersymmetry was originally introduced to solve the hierarchy problem and therefore
fix the electroweak (EW) scale  in the presence of quantum corrections.
Numerical methods testing the SUSY models
often  report a  good  likelihood $L$ (or $\chi^2=-2\ln L$) to
fit the data {\it including} the EW scale itself ($m_Z^0$) with a
{\it simultaneously} large fine-tuning i.e. a large variation of this scale
under a small variation of the SUSY parameters.  
We argue that this is inconsistent and we identify the origin of this problem.
Our claim is that the likelihood (or $\chi^2$) to fit the data 
that is usually reported in such models 
does not  account for the $\chi^2$ cost of  fixing the EW scale.
When this constraint is  implemented, 
the likelihood (or $\chi^2$) receives a significant correction ($\delta\chi^2$)
that worsens the current data fits of SUSY models. We estimate this correction for 
the models: constrained MSSM (CMSSM), models with non-universal gaugino masses (NUGM)
or higgs soft masses  (NUHM1, NUHM2), the  NMSSM and the general NMSSM (GNMSSM).
For a higgs mass $m_h\approx 126$ GeV,  one finds that
in these models  $\delta\chi^2/n_{df}\! \geq  1.5$ ($\approx 1$ for GNMSSM),
which violates the usual condition  
of a good fit (total $\chi^2/n_{df}\approx 1$) 
already {\it before} fitting   observables other than the EW scale itself
($n_{df}$=number of degrees  of freedom). This has  (negative) implications 
for SUSY  models and it  is suggested  that future data fits properly 
account for this effect, if one remains true to the original goal of SUSY.
 Since the expression of $\delta\chi^2$ that emerges 
from our calculation depends on a familiar measure of fine-tuning, 
one concludes that fine-tuning is an intrinsic part of the likelihood 
to fit the data that includes the EW scale ($m_Z^0$).
\end{abstract} 

\newpage

\section{A correction to the likelihood from fixing the EW scale.}

The main  motivation for introducing  low-energy (TeV-scale) supersymmetry (SUSY) 
was to  solve  the hierarchy problem  and therefore  fix the electroweak (EW) scale 
which remains stable under the addition of quantum corrections up to the 
Planck scale.  This is achieved  without a dramatic tuning of the parameters
specific to the  Standard Model (SM)  \cite{Susskind:1978ms}.
SUSY is now under intense scrutiny at the LHC and 
the recent Higgs-like  particle discovery of mass $m_h$ near $126$ GeV \cite{SMH} 
brings valuable  information  on the physics beyond the SM.
In this work we consider  popular SUSY models
and investigate  the  impact on their  data fits of 
the requirement of fixing the EW scale (vev $v$ or $m_Z$),
that SUSY was  supposed to enforce, for the  recently  measured value of $m_h$.

Current  numerical methods that 
test SUSY models against experimental data
involve a remarkable amount of  technical expertise and work, backed by impressive 
computing power, see  for example 
\cite{Kowalska:2012gs,Bechtle:2012zk,Cabrera:2012vu,Fowlie:2012im,Strege:2012bt}.
Somewhat surprisingly,  in the case of a ``frequentist'' approach that we discuss in this work,
there seems to be a  problem  that is overlooked
when computing the likelihood $L$ (or $\chi^2\equiv -2\ln L$) to fit the data.
Let us detail. Numerical methods sometimes report a good likelihood 
(or $\chi^2$) to fit the data\footnote{
A  good  fit should have 
$\chi^2/n_{df}\approx 1, {\rm where}\, n_{df}\equiv n_{\cO}-n_p$, where
$n_{\cO}$ is the number of observables fitted and $n_p$ is the number of parameters.
For a set of observables $\cO_i$ one has total $\chi^2=\sum_i (\cO_i^{th}-\cO_i^{exp})^2/\sigma_i^2$.}
 that  {\it includes} the EW scale itself 
(mass of Z boson, $m_Z^0$), and also a simultaneous 
large\footnote{In the following we do not use a particular mathematical 
definition of fine-tuning.} fine 
tuning\footnote{In some studies $m_Z$
is an {\it input}, so fixing the EW scale is expected to
be respected, but the argument based on eq.(\ref{eq1}) suggests otherwise.
More details are given in Section~\ref{sec3} and  eq.(\ref{puzzle}).},
i.e. a large variation of this scale under a small, fixed
variation of SUSY parameters ($\gamma_i$) of the  
model.  This suggests an inconsistency. To see this, 
 consider a Taylor expansion of the theoretical value of
 $m_Z$ about its very well measured value ($m_Z^0$):
\medskip
\bea\label{eq1}
m_Z=m_Z^0+\Big(\frac{\partial m_Z}{\partial \gamma_i}\Big)_{\gamma_i=\gamma_i^0}\,
(\gamma_i-\gamma_i^0)+\cdots
\eea

\medskip\noindent
The values $\gamma_i^0$ correspond  to the EW ground state, with
$\gamma_i$ the SUSY parameters that define the model, such as 
$m_0$, $m_{1/2}$, $\mu_0$, $A_0$, $B_0$, $\cdots$, in a standard notation. 
A large fine-tuning means  a large partial derivative, then $m_Z-m_Z^0$ suffers a large variation
and that impacts significantly on the value of total $\chi^2$.
So one should expect a poor fit 
in the models with large fine-tuning, however a good fit $\chi^2/n_{df}\approx 1$
 is often reported  in these models\footnote{As we prove later in the text, eq.(\ref{ud}),
the variation of $m_Z$ corresponding to an
EW fine-tuning  of $\cO(1000)$ that is  found in most models, is well
beyond a $\pm 2\sigma$  variation  (around central $m_Z^0$) that is usually taken 
for an input or fitted observable and that actually corresponds to  fine-tuning  $<\!10$.}.
  How do we clarify this puzzle?
($n_{df}$= number of the degrees of freedom).

Our claim, that answers this issue, 
is that in such cases the currently reported likelihood to fit the data (or its $\chi^2$)
does {\it  not} account  for the $\chi^2$ ``cost'' of the condition of
  fixing the EW scale to its  well-measured value
(and that motivated the initial idea of supersymmetry). This leads to an
 underestimate of the overall  value of $\chi^2/n_{df}$ in all  popular 
models used at the LHC. In this work we evaluate the induced correction to $\chi^2/n_{df}$ 
due to this condition, for popular SUSY models. 
The models analyzed  include MSSM-like models with different boundary conditions for  gaugino 
and higgs soft masses,   NMSSM and a generalized version of it, the so-called GNMSSM.

To these purposes, we use the approach of  \cite{DG} which we extend
 to apply beyond its original  restrictive setup that calculated only 
the {\it integrated}  likelihood over the nuisance variables $(y)$. Instead, it is
 the likelihood itself that is required for a traditional, conservative
 frequentist approach used here. We generalize this method so that 
our calculation of the total likelihood function (or $\chi^2$)
keeps explicit its dependence on the nuisance variables such as Yukawa 
couplings. This is important since:  a) it allows one to 
subsequently maximize (profile) the likelihood wrt these variables, which is 
needed in  numerical  applications, and  b)
it avoids the likelihood dependence on the (chosen) measure under 
the integral over nuisance variables.

We show that in all SUSY models,
the likelihood (or  $\chi^2$) receives a correction  due to the condition of  fixing
the EW scale to the measured value ($m_Z^0$) that worsens significantly
 the currently reported $\chi^2/n_{df}$.
We then estimate this correction, hereafter denoted $\delta\chi^2$,
by using the numerical results of 
\cite{Kowalska:2012gs,gz,Cassel:2010px,Ross:2012nr} for a fine 
tuning measure (denoted $\Delta_q$) that emerges in  our calculation of 
$\delta\chi^2\propto \ln \Delta_q$ (see later).
Note that the  value of $\delta\chi^2$ depends strongly   
on the value of the higgs mass $m_h$.  
Using  the recent LHC result  $m_h\approx 126$ GeV,
we find in the models other than the GNMSSM, a  
correction  $\delta\chi^2/n_{df}\!>\!1.5$ 
without including the usual  $\chi^2$ cost due to fixing 
observables other than the EW scale itself.  Therefore,
under the assumption of a simultaneous minimization of both $\delta\chi^2$
and the ``usual'' $\chi^2$ (assumed to respect  $\chi^2/n_{df}\approx 1$) 
these models have a total $(\chi^2+\delta\chi^2)/n_{df}>2.5$, hardly  compatible 
with the data (in the GNMSSM $(\chi^2+\delta\chi^2)/n_{df}\approx 2$).

Given its implications for these models and for SUSY in general,
it is thus suggested that this correction be 
included in future EW data fits that compute  $\chi^2/n_{df}$.
In the following  we substantiate these  claims and analyze the   
consequences for the viability of SUSY models.

\section{The calculation of the  correction to $\chi^2$.}

Let us briefly review the numerical calculation of the likelihood 
in a SUSY model (see for example \cite{BA,Ellis:2007fu}). One usually chooses 
 a set of observables $\cO_i$ well measured 
such as: the W-boson mass,  the  effective leptonic weak mixing angle $\theta_{eff}^{lep}$,
the total Z-boson decay width, the anomalous magnetic moment of the muon, the 
mass of the higgs  ($m_h$), the dark matter relic density, the branching 
ratios from B-physics,  $B_s\!-\!\overline B_s$ mass difference and also additional 
bounds (not shown)  which should be counted in $n_{df}$, too:
\bea\label{EWC}
&& \qquad\qquad
m_W,\,\, \sin^2\theta_{eff}^{lep},\,\, \Gamma_Z,\,\,\delta a_\mu,\,\,m_h,\, \Omega_{DM} h^2,\qquad
\nonumber\\[7pt]
&& BR(B\ra X_s\gamma),\,\, 
BR(B_s\ra \mu^+\mu^-),\,\, 
BR(B_u\ra \tau \nu),\,\,
\Delta M_{B_s}, \,{\rm etc.}
\eea

\medskip\noindent
Note that fixing the EW scale to its accurately measured value ($m_Z^0$)
is not on this list, even though this motivated SUSY in the first place.
However, data fits often include $m_Z^0$ as an {\it input} 
observable, so one could argue  that this observable is indeed being fixed. 
We return to this  issue later in the text\footnote{See  discussion
near eq.(\ref{puzzle}).}.
For each observable $\cO_i$ the corresponding probability
is often taken a Gaussian $P(\cO_i\vert \gamma, y)$
where by $\gamma$ we denote the set of SUSY parameters that define the model,
while by $y$ we denote nuisance variables  such as  Yukawa 
(of top, bottom, etc) and other similar couplings. 
One then assumes that the observables are independent and multiplies 
their probability distributions to obtain a total  distribution, which regarded
as a function of $\gamma$, $y$ 
(with ``data'' fixed), defines the likelihood $L$:
\medskip
\bea\label{def1}
L
\!=\!\prod_j P(\cO_j\vert \gamma, y);
\,\,\,\,\,\,
 \gamma\!=\!
\{m_0,m_{1/2},\mu_0, m_0, A_0, B_0,\cdots\};
\,\,\,\,\,\,
y\!=\!\{y_t, y_b, \cdots\}
\eea

\medskip\noindent
in a standard notation for the SUSY parameters, that are components of the set  $\gamma$.

To work with dimensionless parameters, all $\gamma$ should be ``normalized'' to some scale (e.g. 
 the EW scale vev $v_0\equiv 246$ GeV).
In the ``frequentist'' approach one maximizes   $L$ or equivalently minimizes the value of
$\chi^2$ that, under a common assumption of  Gaussian distributions, is defined as 
\medskip
\bea\label{ihc}
\chi^2=-2\ln L.
\eea 

\medskip\noindent
One then seeks  a good fit, such  that $\chi^2/n_{df}\approx 1$ at the minimum, by 
 tuning $\gamma$, $y$ to fit observables $\cO_i$. 
Fixing the EW scale ($m_Z$) to its measured value ($m_Z^0$)
is not something optional in our opinion, but an intrinsic part of the 
likelihood  to fit the EW data, and in the following we  evaluate
the corresponding $\chi^2$ ``cost'' of doing so.

In other approaches like the ``Bayesian'' method
one further combines $L$ with priors (probabilities for $\gamma, y$)
to obtain the posterior probability  \cite{gz,BA,Allanach:2007qk,Cabrera:2008tj}. 
In this method one searches  for the point with  the largest probability 
in   parameter space, given the data.

While the observables $\cO_i$ are  independent, the SUSY set of  
 parameters $\gamma$ 
used to fit them, are not. They are usually constrained (correlated)
by the EW minimum conditions. 
Indeed, in MSSM-like  models, there are two minimum conditions  of the scalar
potential\footnote{There is an extra condition in the case of NMSSM and GNMSSM models.}; 
one of them is determining the EW 
scale $v$ as a {\it function} of the parameters $\gamma$ of the model
(it is not fixing $v$ to any numerical value).
In practice however, when doing data fits, one usually replaces $v$ 
{\it by hand},  by the measured mass of Z boson ($m_Z^0$) and solves 
this minimum condition for a  SUSY parameter instead (usually $\mu_0$). 
Doing so can miss the impact on $L$ of  the relation between the 
distribution fixing the observable $m_Z\propto v$ to 
its measured  $m_Z^0$, and the distribution fixing $\mu_0$. This is discussed shortly.
The second minimum condition is fixing one parameter (say $\tan\beta$)
as a  function of the remaining\footnote{One can  choose another
parameter (instead of $\tan\beta$) 
like $B_0$, but the effect is just a  change of variables.} $\gamma$.

Let us then consider the scalar potential
for MSSM-like models and evaluate the total likelihood,
hereafter denoted $L_w(\gamma,y$). 
The potential has the following standard expression:
\medskip
\bea
V\!\!&=&\!\!  m_1^2 \vert H_1\vert^2
+ \! m_2^2 \vert H_2\vert^2-\! 
(m_3^2 H_1. H_2+ h.c.)+(\lambda_1/2) \vert H_1\vert^4
\!+(\lambda_2/2) \vert H_2\vert^4
\!+\lambda_3  \vert H_1\vert^2 \vert H_2\vert^2 
\nonumber\\[2pt]
& +&\!\lambda_4 \vert H_1. H_2 \vert^2
 +\big[(\lambda_5/2) (H_1. H_2)^2
 +\lambda_6 \vert H_1\vert^2 (H_1. H_2)+
\lambda_7 \vert H_2 \vert^2 (H_1. H_2)+h.c.\big]
\eea

\medskip\noindent
We denote
\medskip
\bea
\lambda& \equiv &\lambda_1/2\,\cos^4 \beta
+\lambda_2/2 \,\sin^4\beta+(\lambda_3+\lambda_4+\lambda_5)/4\,\sin^2 2\beta
+(\lambda_6\,\cos^2\beta+\lambda_7\,\sin^2\beta);
\nonumber\\[2pt]
m^2 &\equiv  & m_1^2\cos^2\beta\!+m_2^2\sin^2\beta\!-\! m_3^2\sin 2\beta,
\eea

\medskip\noindent
 $\lambda$ is the {\it effective} quartic higgs 
coupling and $m^2$ is a combination of the higgs soft terms\footnote{
The EW min conditions are $v^2=-m^2/\lambda$,
$2\lambda (\partial m^2/\partial \beta)\!-\!m^2(\partial \lambda/\partial \beta)\!=\!0$.
See also \cite{DG,Cassel:2010px}.}.

To compute the total likelihood  $L_w(\gamma,y)$
that also accounts for the effect of fixing the EW scale,
we use the original idea in \cite{DG} which we extend to a general case. 
The result in \cite{DG} is too restrictive since it only provided the
{\it integral} of  $L_w$ over nuisance variables like the 
Yukawa couplings $(y)$,  giving $L(\gamma)\sim \int d y L(\gamma,y)$.
Such integrated likelihood is  
not appropriate for a traditional, conservative ``frequentist'' approach. 
To avoid this shortcoming, we  compute 
 $L_w(\gamma,y)$ itself, rather than its  integral  over the set $y$
 (or corresponding masses). This  allows one 
to subsequently {\it maximize} $L_w(\gamma,y)$ with respect to 
 nuisance  variables  for a fixed set of $\gamma$ parameters (to find
the ``profile likelihood'').

It is useful to write the two EW minimum conditions 
 as Dirac delta of two functions $f_{1,2}$ defined below
\medskip
\bea
f_1 & \equiv& v-(-m^2/\lambda)^{1/2},
\nonumber\\
f_2 &\equiv & \tan\beta-\tan\beta_0(\gamma, y, v),\qquad {\rm where}
\qquad f_i=f_i(\gamma; y, v, \beta), \,\,i=1,2.
\label{min2}
\eea

\medskip\noindent
$\beta_0$ is the root of  the second minimum condition; $\beta_0, f_{1,2}$ depend
on the arguments shown; 
$\lambda$, $m^2$ also depend on $\gamma$, $y$, $\beta$.
Taking account of constraints (\ref{min2}) the total  (``constrained'') 
likelihood $L_w(\gamma,y)$  is
\medskip
\bea
 L_w(\gamma,y)\!\! &=&\!\! 
m_Z^0\! \int  d v\,  d(\tan\beta)\,\, \delta \big[f_1(\gamma; y, v, \beta)\big]\,\,
\delta\big[f_2(\gamma; y, v, \beta)\big]\,
 \delta(m_Z-m_Z^0)\,\, L(\gamma; y, v,\beta)
\nonumber\\[1pt]
\!\!\!&=&\!\!
v_0\,\, 
L\big(\gamma; y, v_0, \beta_0(\gamma,y)\big)
\,\,\delta\big[ f_1\big(\gamma; y, v_0, \beta_0(\gamma,y)\big)\big]
\nonumber\\[5pt]
&=&L(\gamma,y,v_0,\beta_0(\gamma,y))\,\,\delta(1-\tilde v/v_0);
\qquad\qquad \tilde v\equiv (-m^2/\lambda)^{1/2}\Big\vert_{\beta=\beta_0(\gamma,y),v=v_0}.
\label{eqeqp}
\eea

\medskip\noindent
Here $v_0=246$ GeV, $m_Z^0\!=\!g\,v_0/2\approx 91.2$ GeV, $m_Z\!=\!g\,v/2$, $g^2\!=\!g_1^2\!+\!g_2^2$
with $g_{1,2}$ couplings for U(1), SU(2). To simplify notation,
we did not display the numerical argument $v_0$ of  $\beta_0$, i.e.  we denoted
$\beta_0(\gamma,y)\!\equiv\! \beta_0(\gamma,y,v_0)$. $m_Z^0$ in rhs
compensates the dimension of $\delta(m_Z\!-\!m_Z^0)$.

$L(\gamma, y, v, \beta)$ in the first line of  eq.(\ref{eqeqp})
 denotes the usual likelihood associated with
fitting the  observables {\it other than} the EW scale ($m_Z$), see eq.(\ref{EWC}), before
 $v$, $\tan\beta$ are fixed  by  the EW min conditions.
Integrating over $v$, $\tan\beta$ in the  presence of the delta functions
  $\delta(f_1)$, $\delta(f_2)$
is just a formal way to solve these EW minimum constraints and  eliminate
these parameters in terms of the  rest.
We also introduced a $\delta(m_Z-m_Z^0)$ distribution, 
which fixes the EW scale by enforcing the  measured mass of Z boson ($m_Z^0$) and thus the 
replacement $v\ra v_0$.
Since  $m_Z^0$  is very  well measured, using  $\delta(m_Z-m_Z^0)$ is indeed justified
(however, the result can be generalized to a Gaussian\footnote{
Additional distributions  can also be considered  
for nuisance variables, like  top, bottom masses, etc, and assumed to factorize out 
of $L$ in the rhs,   with corresponding integrals over Yukawa  couplings, see \cite{DG}.}).

A comment about normalization in (\ref{eqeqp}): 
one must ensure that the two constraints ($f_{1,2}$)
are indeed  normalized to unity wrt the variables over which we integrate them, in this 
case $\tan\beta$ and $v$ (this condition is indeed respected in our case\footnote{$\delta(f_{1,2})$
are just distributions for $v,\tan\beta$  (albeit special ones), so they also  must 
be normalized to unity.}). Therefore  there is  {\it no freedom}
 for any additional factors to be present in eq.(\ref{eqeqp}).

The last two lines in eq.(\ref{eqeqp})  show that the original likelihood $L$,
evaluated on the ground state,  is now multiplied by
  $v_0\,\delta(f_1(...,v_0,...))=\delta(1-\tilde v/v_0)$.
Further, the Dirac delta of a  function $f_1(z_i)$, $i=1,2,...n$, can be written as
(after a Taylor expansion of $f_1$):
\medskip
\bea\label{hh}
\delta (f_1 (z_i))
=\delta\big[f_1(z_i^0)+ (\nabla f_1)_0. (\vec z-\vec z^0)+\cdots\big]
= \frac{1}{\vert \nabla f_1\vert_0}\,
\delta [n_j\,(z_j- z^0_j)],
\eea

\medskip\noindent
A summation over repeated index $j$ is understood and the
subscript ``o'' of  $\vert \nabla f_1\vert_0$ means this quantity is evaluated
at the point $z_i\!=\!z_i^0$ where $f_1\!=\!0$; $n_i$ are components of the normal $\vec n$
to the surface $f_1(z_i^0)=0$, so  $\vec n=(\nabla f_1/\vert\nabla f_1\vert)_0$. 
Eqs. (\ref{eqeqp}), (\ref{hh}) tell us  that it is not enough 
for the parameters of the model to respect the constraint $f_1=0$,
and that there is instead  an {\it additional}  factor generated, represented by the 
gradient of the constraint\footnote{
The gradient  measures how the constraint ``reacts''
 to the variations of the  parameters.}.

The ``constrained'' likelihood of eq.(\ref{eqeqp}) becomes, after using eq.(\ref{hh}):
\medskip
\bea
L_w(\gamma,y)\!=\!
\frac{\delta\big[n_i (\ln z_i-\ln z^0_i)\big]}{\Delta_q(\gamma^0\!,y^0)}\,
\,
L\big( \gamma; y,\,v_0,\beta_0(\gamma,y)\big),
\qquad  z_i\!=\!\{\gamma_j, y_k\}
\label{oo}
\eea

\medskip\noindent
Here $\gamma_j$, $y_k$ denote components of the sets $\gamma$ and $y$ defined 
in (\ref{def1}) and we used (\ref{hh}) for $\ln z_i$ instead of $z_i$ as variables,
to ensure dimensionless arguments for $\delta$ function.
$\Delta_q$ denotes the absolute value of the gradient of $f_1$ evaluated
at  $z_i^0=\{\gamma^0_j, y^0_k\}$ that is a solution to
the EW min condition  $f_1(\gamma^0,y^0,v_0,\beta_0(\gamma^0,y^0))\!=\!0$, 
and has the value:
\medskip
\bea
\label{deltaq}
\Delta_q^2(\gamma^0\!\!, y^0)=\!\!\!\!\!\!
\sum_{z_i=\{\gamma_j, y_k\}}\!\!\!
 \Big(\frac{\partial \ln \tilde v}{\partial \ln z_i} \Big)^2_{\! o}
\!\!= \sum_{\gamma_j} 
 \Big(\frac{\partial \ln \tilde v}{\partial \ln \gamma_j} \Big)^2_{\! o}
+\!\sum_{y_k} 
\Big(\frac{\partial \ln \tilde v}{\partial \ln y_k} \Big)^2_{\! o}.
\eea

\medskip\noindent
The subscript ``o''  stands for evaluation on the ground state
 ($\gamma_i\!=\!\gamma_i^0$, $y_k\!=\!y_k^0$).
 $\Delta_q$ that emerged above has some resemblance to what 
is called the fine  tuning measure\footnote{This was introduced in \cite{Ellis:1986yg},
with $\Delta_{max}=\max_\gamma \vert \partial \ln\tilde v/\partial\ln \gamma_i\vert$;
see \cite{CS} for an interpretation of $1/\Delta_{max}$.}
wrt all parameters, both $\gamma$ and $y$.
The arguments of $\Delta_q(\gamma^0,y^0)$ denote the parameters wrt which is 
computed and includes SUSY parameters {\it and} nuisance variables (Yukawa, etc).

Eq.(\ref{oo}) is the result in terms of distributions and
can be written in an alternative form.
The  $\delta$  in the rhs of (\ref{oo}) tells us that the lhs is non-zero when
$z_i=z_i^0$ for all $i$ (i.e. $\gamma_j\!=\!\gamma_j^0$, $y_k\!=\!y_k^0$).
Another way to present this eq is to do a formal integration of (\ref{oo}) in 
the general direction  $\ln \tilde z= n_j \ln z_j$,  (sum over $j$ understood, $j$ running
over the sets $\gamma, y$);
this allows all independent parameters to  vary simultaneously\footnote{Such integral is just
a formal way of saying we solve $f_1=0$ in favour of one particular $\gamma^0_\kappa$ (usually 
$\mu_0$).}. After this,  eq.(\ref{oo}) becomes:
\medskip
\bea\label{ttt}
L_w(\gamma^0\!, y^0)=
\frac{L\big(\gamma^0; y^0, v_0, \beta_0(\gamma^0,y^0)\big)}
{\Delta_q(\gamma^0\!\!, y^0)}.
\eea

\medskip\noindent
$L$ in the rhs is exactly the usual, ``old'' likelihood computed in the 
 data fits  and evaluated at the EW minimum (reflected by its arguments)
 but {\it without} fixing the EW scale,
 while the lhs does account for this effect.  
 Note that due to the
minimum condition $f_1(\gamma^0,y^0,v_0,\beta_0(\gamma^0,y^0))\!=\!0$, one
element of the set $\gamma^0$, say $\gamma_\kappa^0$,  becomes a function of 
the remaining, independent $\gamma^0_i$, $i\not=\kappa$. Usually
$\gamma_\kappa^0$ is taken to be $\mu_0$.

The result in eq.(\ref{ttt}) shows that for a good fit (i.e. maximal
 $L_w$),  one has to maximize not the usual likelihood in the rhs, 
but actually its {\it ratio} to $\Delta_q$.
Let us introduce the notation $\chi^2_w\equiv-2\ln L_w$  and also 
$\chi^2\equiv -2\ln L$, then eq.(\ref{ttt}) becomes
\medskip
\bea\label{chisq2}
\chi^2_{w}(\gamma^0,y^0)=\chi^2(\gamma^0,y^0)+2\ln \Delta_q(\gamma^0,y^0).
\eea

\medskip
Therefore, after fixing the EW scale
the usual $\chi^2$ receives a positive correction
that depends on $\Delta_q$ and that is not included 
in the current precision
data fits. This result extends the validity of  its counterpart in  \cite{DG},
 {\it in the presence} of the nuisance variables $y$. Unlike in \cite{DG},
nuisance variables $y$ are present  in eq.(\ref{chisq2}) i.e. were not
integrated out. This has the 
 advantage of respecting the traditional, conservative frequentist approach
and  one can compute numerically from eq.(\ref{ttt}) the profile 
likelihood,  by maximizing $L_w$   wrt
nuisance variables $y^0$, for a fixed  set of $\gamma^0$. This profile likelihood  
is then $L_w(\gamma^0,y^0_{max}(\gamma^0))$.

Let us remark that in eq.(\ref{oo})
 we used  $\ln z_i$ as arguments under the Dirac delta function,
which implicitly assumes that these are more fundamental parameters than $z_i$ themselves,
(here $z_i=\{\gamma_j,y_k\}$). In principle this is a choice,  motivated here 
by the fact that it  ensured  dimensionless arguments for the delta function in 
(\ref{oo}), (unlike $y_k$, $\gamma_j$ are dimensionful parameters).
Going from  these parameters to their log's
is a one-to-one change that  does not affect the minimal value of
total\footnote{In a Bayesian language,
 this would correspond to chosing log priors instead of flat ones
for the parameters. Ultimately this may reflect a problem of {\it  measure}
that is beyond the purpose of this work.} $\chi_w^2$.
If one insists in working with $z_i$  as fundamental parameters,
one simply changes  $\ln z_i\ra z_i$ in eqs.(\ref{oo}), (\ref{deltaq}), after
``normalizing'' $\gamma_i$ to some scale (e.g. $v_0$), to ensure 
dimensionless arguments for $\delta$ in eq.(\ref{oo}). 
The result is that $\Delta_q$  is then computed wrt
$\gamma_j$ and $y_k$ instead of their logarithms, so in eqs.(\ref{ttt}), (\ref{chisq2})
 one  replaces
\medskip
\bea\label{op}
\Delta_q^2\ra \Delta_q^{\prime\, 2}=\sum_{\gamma_j}
 \Big(\frac{\partial \ln \tilde v}{\partial \gamma_i}\Big)_{\! o}^2
+\sum_{y_k}\Big(\frac{\partial \ln \tilde v}{\partial y_k}\Big)_{\! o}^2
\eea

\medskip\noindent
Compared to its counterpart in eq.(\ref{deltaq}),
the second sum above is actually larger in this case since $y_k<1$.
In the following, for numerical estimates  we shall work with  $\Delta_q$.
A detailed investigation  of the correction to $\chi^2$
is beyond the purpose of this paper and in the following we restrict the study
to a numerical estimate. The main point is 
that $\chi^2$ has a correction that needs to be taken into 
account.

As argued in the introduction, numerical methods to fit the data seem to
 miss the above effect, leading to the puzzle  mentioned near
 eq.(\ref{eq1}), of having simultaneously a good fit
  $\chi^2/n_{df}\approx 1$ but a  large fine-tuning (that should actually worsen this fit!).
Usually in the numerical methods one often uses $m_Z=m_Z^0$ as an input, i.e. with a 
Dirac delta distribution,  to compute instead the corresponding value 
$\mu_0$, which itself has a similar distribution; there is however
 a relative normalization  factor between these, 
that is relevant in answering this issue.
To see this, consider that all parameters $\gamma_i, y_k$, $i\not=1$ 
are fixed to some numerical values  ($\gamma_i^0$, $y_k^0$), except $\gamma_1\equiv \mu_0$. 
Then\footnote{under the assumption of a unique root for $\mu$.}
\bea\label{puzzle}
L_w=L\,\delta(1-\tilde v/v_0)=L \,\,\delta(1-m_Z/m_Z^0) = L \,\frac{ \delta (1-\mu/\mu_0)}{\Big\vert
 \frac{\partial \ln m_z}{\partial\ln \mu}\Big\vert_0}
\eea

\medskip\noindent
The denominator  is a particular version  of our more 
general $\Delta_q$ when only one parameter varies, and is
just a normalization factor. When missing this factor, 
the likelihood  of the model does not account for the ``cost'' of 
fixing the EW scale (i.e. $m_Z^0$). 
This  factor  affects the minimal value of $\chi^2$ and must be included in total
$\chi^2/n_{df}$, as shown in eq.(\ref{chisq2}) when all parameters vary.
The above discussion answers the  puzzle 
and clarifies how a numerical method can be adapted to
account for the $\chi^2$ ``cost'' of fixing the 
EW scale. The  discussion can be extended to
more  general likelihood functions (beyond Dirac delta type).

The ``frequentist'' approach shown so far 
can be extended  to the Bayesian case which 
is just a global version (in parameter space) of eq.(\ref{ttt}).
In this case 
one assigns, in addition,  some initial probabilities (priors) to the parameters 
of the model ($\gamma, y$) then integrates over them
 the  likelihood $L_w$ multiplied by the priors. This gives
 the global probability of the model or ``evidence'', $p({\rm data})$, that must be
maximized. The result is (see \cite{gz,Fichet}):
\bea
p({\rm data})=\int_{f_1=f_2=0} \!\!\!
 d S \,\frac{1}{\Delta_q(\gamma, y)}\,\,L(\gamma, y, v_0, \beta)
\,\times\,{\rm priors}(\gamma,y)
\eea

\medskip\noindent
where the  integral is over a surface in the parameter space $\gamma, y$ 
defined by $f_1=f_2=0$.
In this case the factor  $1/\Delta_q$ is itself an ``emergent'', naturalness prior, 
in addition to  the original priors of the model.
Unlike the frequentist approach, the effect of fixing the EW scale
(associated with $1/\Delta_q$)
is indeed taken into account in the Bayesian approach \cite{Allanach:2007qk,Cabrera:2008tj}
(under some approximation). For details on the Bayesian case, 
 see   \cite{BA,gz,Allanach:2007qk,Cabrera:2008tj,Fichet,Delta,Balazs:2012qc}.

\section{A numerical estimate of the correction  to  $\chi^2$ in SUSY.}
\label{sec3}

 In this section we estimate in some models the  correction $\delta\chi^2$ 
obtained from eq.(\ref{chisq2})
\bea\label{dchi}
{\delta\chi^2}/{n_{df}}\equiv (2/n_{df})\,\ln\Delta_q
\eea

\medskip\noindent
By demanding that the correction $\delta\chi^2/n_{df}$ be small
enough not to affect the current data fits results for $\chi^2/n_{df}$, 
one has a model-independent upper bound:
\bea
\Delta_q \ll \,\exp\big(n_{df}/2\big).
\eea
When this bound is reached, then $\chi^2_w/n_{df}=1+\chi^2/n_{df}$,
and assuming  values of ``usual'' $\chi^2/n_{df}\approx 1$ i.e. a good fit 
without fixing the EW scale, gives $\chi^2_w/n_{df}\approx 2$.

Using (\ref{dchi}), one can find the equivalent numerical
value of $\Delta_q$  that  corresponds to one observable having a 
given deviation from the  central value. 
This can be read below:
\bea\label{ud}
2\,\sigma  &\leftrightarrow & \Delta_q\approx 8.\nonumber\\
3\,\sigma  &\leftrightarrow & \Delta_q\approx 100.\nonumber\\
3.5\,\sigma  & \leftrightarrow  & \Delta_q\approx 1000.\nonumber\\
5\, \sigma   &  \leftrightarrow  & \Delta_q\approx 1\,000\, 000.
\eea

This gives, in a model independent way, a different perspective and a
probabilistic interpretation to the  values of  $\Delta_q$ 
(related to fine-tuning/naturalness) that avoid subjective criteria
about this topic.  As a result, any model with $\Delta_q\!>\!100$  ($\Delta_q>200$) 
such as those discussed shortly,
is more than a $3\sigma$ ($3.25 \sigma$) away
from fixing the EW scale ($m_Z$). This supports our initial discussion 
(near eq.(\ref{eq1})), that   one cannot have a 
good fit of $m_Z$ with a simultaneous, large EW fine tuning
(assuming $m_Z$ is independent of  other observables).

For our numerical estimates of $\delta\chi^2$ 
we restrict the analysis to using  {\it minimal} values of $\Delta_q$ in  SUSY 
 models.  Further, we  only evaluate $\Delta_q$ wrt $\gamma$
parameters; notice that
\bea
\Delta_q(\gamma, y)>\Delta_q(\gamma).
\eea
\noindent
where the arguments are the variables wrt which $\Delta_q$ is actually computed.
 That is, we shall ignore the contribution
to $\Delta_q$ due to variations  wrt Yukawa couplings or other nuisance parameters.
This  underestimates  $\delta\chi^2$, but has the advantage that 
we can use the  numerical values of $\Delta_q$  already available,
in \cite{gz} (also \cite{Cassel:2010px}) for MSSM models
 with different boundary  conditions (two-loop results) and in
\cite{Kowalska:2012gs,Ross:2012nr} for NMSSM, GNMSSM  (one loop).

 We  consider the most popular SUSY models  
used for searches at the  LHC, listed  below:

\noindent
$\bullet$ the constrained MSSM model (CMSSM); this is the basic scenario, 
of parameters\footnote{Parameters $\gamma$ are those
 following from (\ref{oo}), (\ref{deltaq}) and this is why $\mu_0$ is quoted 
instead of usual $sgn(\mu_0)$. Similar for the other models. 
Also $B_0$ is quoted instead of $\tan\beta$, 
the difference is a change of variables. }
 $\gamma \equiv \{m_0, m_{1/2}, \mu_0, A_0, B_0\}$, in a standard
notation.
Then $\Delta_q$ is that of  eq.(\ref{deltaq}) with summation over these
 parameters only:
\bea\label{dq}
\Delta_q^2=\sum_{\gamma_j} 
\Big(\frac{\partial \ln \tilde v}{\partial \ln \gamma_j}\Big)^2_{\! o}
\eea

\medskip
\noindent
$\bullet$  the NUHM1 model:  this is a CMSSM-like model in which one relaxes
the   Higgs soft  masses in the ultraviolet (uv), to allow values 
different from $m_0$: $m_{h_1}^{uv}=m_{h_2}^{uv}\not= m_0$, with
$\gamma\equiv \{m_0, m_{1/2}, \mu_0, A_0, B_0, m_{h_1}^{uv} \}$.
$\Delta_q$ is as in eq.(\ref{dq}) with summation over this set.

\medskip
\noindent
$\bullet$ 
the NUHM2 model: this is a CMSSM-like model with non-universal Higgs soft 
masses,  $m_{h_1}^{uv}\!\not=\!m_{h_2}^{uv}\!\not=\! m_0$,  
with independent parameters  $\gamma\equiv 
\{m_0, m_{1/2}, \mu_0, A_0, B_0, m_{h_1}^{uv},m_{h_2}^{uv}\}$.
Then  $\Delta_q$ is that of eq.(\ref{dq})  with summation over this set.

\medskip
\noindent
$\bullet$ the NUGM model:
 this is a CMSSM-like model with non-universal gaugino masses 
$m_{\lambda_i}$, $i=1,2,3$,  with 
$\gamma=\{m_0, \mu_0, A_0, B_0, m_{\lambda_1}$, $m_{\lambda_2}, m_{\lambda_3}\}$.
 $\Delta_q$ is that of eq.(\ref{dq}) with summation over this set.

\medskip
\noindent
$\bullet$ the NUGMd model: this is a particular case of 
the NUGM  model with a specific relation among 
the gaugino masses $m_{\lambda_i}$, $i=1,2,3$,  of the type 
$m_{\lambda_i}=\eta_i\, m_{1/2}$,
where $\eta_{1,2,3}$ take only {\it discrete}, fixed values. Such relations can exist
due to some GUT symmetries, like SU(5), SO(10), etc \cite{Horton:2009ed}.
The particular relation we consider is a benchmark point  with 
$m_{\lambda_3}=(1/3) \,m_{1/2}$, 
$m_{\lambda_1}=(-5/3)\, m_{1/2}$, 
$m_{\lambda_2}=m_{1/2}$, corresponding to a particular GUT (SU(5)) model, 
see Table~2 in \cite{Horton:2009ed}.
$\Delta_q$ is that of eq.(\ref{dq}) with 
$\gamma=\{m_0, m_{1/2}, A_0, B_0, \mu_0\}$.

\medskip
\noindent
$\bullet$ the next to minimal MSSM model (NMSSM): the model
has an additional gauge singlet. Its parameters are
$\gamma=\{m_0, \mu_0, A_0, B_0, m_{1/2}, m_S\}$, with $m_S$ the singlet soft mass.
For an estimate of $\Delta_q$ we use the results in 
\cite{Kowalska:2012gs,Ross:2012nr}. Notice that these papers evaluate
instead $\Delta_{max}$ which is the largest fine tuning wrt to any of these parameters
(instead of their sum in ``quadrature'' as in $\Delta_q$). As a result, $\Delta_{max}$ is usually
slightly smaller, by a factor between 1 and 2 as noticed
for the other models listed above \cite{gz} and thus 
$\delta\chi^2$ is underestimated.

\medskip
\noindent
$\bullet$ the general NMSSM (GNMSSM) model: this  is an extension  of the NMSSM in 
the sense that it contains a bilinear term
in the superpotential for the gauge singlet, $M S^2$, see for example 
\cite{Ross:2012nr}. So the singlet is massive at the supersymmetric 
level\footnote{Its  mass can be of few  (5-8) TeV, so
one can integrate it out and work near the decoupling limit \cite{Cassel:2009ps}.}
and we have an additional parameter $M$ to the set we have for the NMSSM.
Again, $\Delta_{max}$ is used here \cite{Ross:2012nr} instead of 
$\Delta_q$, so $\delta\chi^2$ is again underestimated.

Our estimates for $\delta\chi^2/n_{df}$ 
for different higgs mass values
are shown in Tables~\ref{tabel1} and \ref{tabel2}.
The results present the   {\it minimal} value of 
$\Delta_q$ evaluated in the above models as a function of the higgs mass,
after a scan over the entire  parameter space (all $\gamma$, $y$ and also $\tan\beta$
of the corresponding model, as described in\footnote{The scan included a range of 
$[-7,7]$ TeV for $A_0$, $m_0$ and  $m_{1/2}$ up to 5 TeV, and 
$2\leq \tan\beta\leq 62$ and also  allowed a  $2\sigma$ deviation for 
fitted observables and for a $3\sigma$ for $\Omega_{DM}h^2$, see for details \cite{gz}.  }  \cite{gz}).
We show the values of $\Delta_q$ for central values of $m_h$ close to the 
pre-LHC lower bound $\approx 115$ GeV in Table~\ref{tabel1}
  and for $123$ to $127$ GeV in Table~\ref{tabel2}.
These values allow us to account for a 2-3 GeV error of the 
theoretical calculation at 2-loop leading log level \cite{theor-higgs-err}.
For $m_h\approx 115$ GeV, one could still have  $\delta\chi^2/n_{df}<1$ 
for some models, that could have allowed  a corrected
$\chi^2_w/n_{df}\approx 1$.
Further, minimal $\Delta_q$ grows approximately  exponentially wrt to $m_h$, due 
to quantum corrections. Indeed,  $\Delta_q\sim m_{susy}^2$ and since
the loop correction is roughly $\delta m_h\sim \ln m_{Susy}$ one finds
 $\Delta_q\approx \exp(m_h/{\rm GeV})$.
As a result, a strong variation of $\Delta_q$ wrt $m_h$ is found, 
 and   $\delta\chi^2$   increases by 
$\approx 1$   for a 1 GeV increase of $m_h$, see Table~\ref{tabel2}.

\begin{table}[!t]
\centering
\begin{tabular}{lclrcrlrlrl}
\hline\hline\\[-7pt]
Model  &  $n_p$  &  Approx  & $\Delta_q$ &  $\delta\chi^2${\footnotesize (115)}  &  $n_{df}$ \\[5pt]
\hline\\[-7pt]
CMSSM  &    5      &  2-loop  &    15  &  5.42  &  9  \\[0.5ex]

NUHM1  &    6      &  2-loop  &   100  &  9.21  &  8  \\[0.5ex]

NUHM2  &    7      &  2-loop  &    85  &  8.89  &  7 \\[0.5ex]

NUGM   &    7      &  2-loop  &    15  &  5.42  &  7 \\[0.5ex]

NUGMd  &    5      &  2-loop  &    12  &  4.97  &  9 \\[0.5ex]

NMSSM  &    6      &  1-loop  &    12  &  4.97  &  8 \\[0.5ex]

GNMSSM &    7      &  1-loop  &     12 &  4.97  &  7 \\[1ex]
\hline 
\end{tabular}
\\[1ex]
\caption{\small
The correction $\delta\chi^2\equiv 2\ln\Delta_q(\gamma)$,
 the number of parameters $n_p$ and degrees of freedom $n_{df}$ in  SUSY
 models, for $m_h\approx 115$ GeV corresponding to the
 pre-LHC  bound of $m_h$. Notice that in this case $\delta\chi^2/n_{df}<1$. 
 $n_{df}$ may vary, depending on the exact number of observables fitted.
The numerical values of $\Delta_q$ are from \cite{gz}
for the first 5 models and  \cite{Ross:2012nr} for NMSSM, GNMSSM
(see also \cite{Cassel:2010px} for CMSSM).}
\vspace{0.3cm}
\label{tabel1}
\end{table}
\begin{table}[!ht]
\centering 
\begin{tabular}{lrlrlrlrl}
\hline\hline\\[-7pt] Model  &   
$\Delta_q$  &  $\!\delta\chi^2${\footnotesize (123);} &
$\Delta_q$  &  $\!\delta\chi^2${\footnotesize (125);} &
$\Delta_q$  &  $\!\delta\chi^2${\footnotesize (126);} & 
$\Delta_q$  &  $\!\delta\chi^2${\footnotesize (127);} \\[5pt] \hline\\[-7pt]
CMSSM &    380 &  11.88 & 1100 & 14.01 & 1800 & 14.99 & 3100 & 16.08 \\[0.5ex]

NUHM1 &    500 & 12.43 & 1000 & 13.82 & 1500 & 14.63 & 2100 & 15.29 \\[0.5ex]

NUHM2 &    470 & 12.31 & 1000 & 13.82 & 1300 & 14.34 & 2000 & 15.20 \\[0.5ex]

NUGM  &    230 & 10.88 & 700  & 13.10 & 1000 & 13.82 & 1300 & 14.34 \\[0.5ex]

NUGMd &    200 & 10.59 & 530  & 12.55 & 850  & 13.49 & 1300 & 14.34 \\[0.5ex]

NMSSM &    $>$100 &  9.21 & $>$200  & 10.59 & $>$200  &  10.59  &  $>$200  &  10.59  \\[0.5ex]

GNMSSM & 22 &  6.18 &   25  & 6.43 & 27 & 6.59 & 31  & 6.87\\[1ex]
\hline 
\end{tabular}
\\[1ex]
\caption{\small
As for Table 1, with $\Delta_q$ and corresponding $\delta\chi^2\equiv 2\ln\Delta_q(\gamma)$  for
$m_h$ equal to 123, 125, 126 and 127 GeV (shown within brackets).
This can also show the impact of the 2-3 GeV error in  the  theoretical calculation of $m_h$
\cite{theor-higgs-err,Degrassi:2002fi}.
 $\Delta_q$ grows $\approx$ exponentially with $m_h$ \cite{gz,Cassel:2010px}; 
 a 1 GeV increase of $m_h$ induces about 1 unit increase  of $\delta\chi^2$.
In all cases except GNMSSM, fixing the EW scale brings a correction
 $\delta\chi^2/n_{df}>1$. 
Note $\delta\chi^2/n_{df}$ can be larger if one also includes the impact of 
Yukawa couplings on $\Delta_q$.  Same loop approximation,
number of parameters and degrees of freedom apply as in Table~\ref{tabel1}.
The  values of $\Delta_q$ are from \cite{gz} for the first five models
and from  \cite{Ross:2012nr} for NMSSM, GNMSSM (see also \cite{Cassel:2010px} for CMSSM).}
\label{tabel2}
\end{table}

\begin{table}[!ht]
\centering 
\begin{tabular}{ccccccc}
 \hline\hline\\[-7pt] 
 $m_h\!=\!126$\,GeV:   & {\rm CMSSM} & {\rm  NUHM1} &  NUHM2 &  NUGM & NUGMd &  GNMSSM\nonumber\\
$\chi^2_w/n_{df}:$  & 2.66; &   2.83;  &  3.05;  &  2.97; &  2.49; &  1.94\\[1ex]
 \hline
\end{tabular}
 \\[1ex]
\caption{\small An estimate for total $\chi^2_w/n_{df}$ in various 
SUSY  models for $m_h\!\approx\! 126$ GeV and $\chi^2/n_{df}\!\approx\! 1$.}
\label{tabel3}
\end{table}

In all models, for the currently measured 
 $m_h\approx 126$ GeV, $\delta\chi^2/n_{df}$ alone is 
larger than unity (or  close to 1 for GNMSSM), 
{\it without} considering the original contribution due to the
``usual'' $\chi^2/n_{df}$ coming from fitting  observables
other than the EW scale. 
The above correction is too large  by the usual criteria 
that total $\chi^2_w/n_{df}\approx 1$. 
Further,  assume that in the above  models one finds a point in the
parameter space for which $\chi^2/n_{df}\approx 1$ and then
add to it the effect of $\delta\chi^2$. One then finds that
 $\chi^2_w/n_{df}$ is close to or larger than 2, see Table~\ref{tabel3}.
In this Table, the values of reduced $\chi^2$  increase (decrease) by $\approx 0.1$ for
an increase (decrease) of $m_h$ by 1~GeV, 
respectively, except in the GNMSSM  where this  is even  smaller (0.04 for 1~Gev).
The log dependence  $\delta\chi^2\!=\!2\ln\Delta_q$ 
means that uncertainties in evaluating $\Delta_q$ are reduced
and $\delta\chi^2$ values listed in Table~\ref{tabel2} and used in Table~\ref{tabel3}
 are  comparable, even though 
corresponding $\Delta_q$'s are very different.

Our  estimates for $\chi^2_w/n_{df}$  shown in Table~\ref{tabel3}
 are hardly acceptable for a good 
fit\footnote{Assuming a $\chi^2$ distribution, 
the $p$-value in these models would be $< 1\%$ (and $5\%$ for GNMSSM).}.
 Interestingly,  increasing the number of parameters of a model 
(decrease $n_{df}$) could  decrease $\Delta_q$ and  $\delta\chi^2$, but
this reduction may not be enough to reduce $\chi^2_w/n_{df}$, since 
$n_{df}$ is now smaller. This is seen by comparing the 
 reduced $\chi^2$ in  NUGM and CMSSM.
This is because $\delta\chi^2$ depends only mildly (log-like) 
on $\Delta_q$,  so only a significant reduction of $\Delta_q$ can compensate
 the effect of simultaneously reducing $n_{df}$. For such case
compare CMSSM with GNMSSM.

The  values of  $\chi^2_w/n_{df}$ could be higher than our estimates above, 
since they ignore that: a) we used only minimal values of $\Delta_q$ 
over the whole parameter space.
b) Yukawa effects on $\delta\chi^2$ were ignored and these can be significant\footnote{
In CMSSM, $\Delta_q$ wrt top Yukawa alone is larger than that wrt to any SUSY
parameter, see fig.2 in \cite{Cassel:2010px}.}. 
There is also an argument in favour of a smaller $\chi^2_w/n_{df}$,
 that current theoretical calculations of $m_h$ may have a 2-3 GeV error.
Assuming this, for  $m_h\approx 123$ GeV (instead of $126$ GeV), in GNMSSM one 
obtains a small change: $\chi^2_w/n_{df}=1.8$  while for the other models this ratio 
is $\geq $2.3. Another reduction may emerge in  numerical analysis
if using eq.(\ref{op}) instead of (\ref{deltaq}), but the impact of its larger 
Yukawa contributions makes this possibility less likely.

Regarding  the NMSSM model, Table~\ref{tabel2} only provided a lower bound on $\Delta_q$. 
However we  can  do a more accurate estimate, 
using recent data fits\footnote{For other  recent data fits see 
\cite{Bechtle:2012zk,Cabrera:2012vu,Fowlie:2012im,Strege:2012bt}.}
that evaluated both $\chi^2$ and $\Delta_q$.
 To this purpose, we  use the minimal  value for $\chi^2\!=\!6.4$ in
 \cite{Kowalska:2012gs} (last two columns of their Table~3)
together with  its corresponding\footnote{
 $\Delta_q$ is even higher as it is not computed according to (\ref{dq})
but reports max values wrt each parameter.}
$\Delta_q\!=\!455$. This gives a $\chi^2_w/n_{df}\!=\!18.64/8$, which is 
similar to the other models discussed above\footnote{
Other values quoted in  \cite{Kowalska:2012gs} bring an even larger value for this ratio.}.
Further, using instead our Table~\ref{tabel2} where
 $\delta\chi^2>10.59$ we find $\chi^2_w/n_{df}>2.32$ 
(for $\chi^2/n_{df}\approx 1$), which is in agreement with the aforementioned
 value derived from accurate data fits.

To conclude, the requirement of fixing the EW scale brings a significant contribution to  
the value of $\chi^2_w/n_{df}$, with negative impact on the
data fits  and on  the phenomenological viability of these models. 
While our numerical results are just an estimate of the correction $\delta\chi^2$,
 the effect is nevertheless present  and demands  a careful re-consideration of this 
correction  by the precision data fits that should  include it in future analysis.

\section{Some implications for model building.}

Let us discuss some implications of the above result for  model building.

\medskip
\noindent
{\bf a).} 
A natural question is how to reduce $\delta\chi^2/n_{df}$. Here
 are  three ways to attempt this:
i) additional  supersymmetric terms in the model, which unlike SUSY breaking ones, 
are less restricted by experimental bounds;
ii) additional gauge symmetry,
iii) additional massive states coupled to the higgs sector.
All these directions
 have in common a possible increase of the {\it  effective} quartic higgs coupling ($\lambda$)
so one can more easily satisfy an EW minimum condition\footnote{
A tension in the relation $v^2=-m^2/\lambda$ 
translates into a larger $\delta\chi^2\sim \ln\Delta_q$ \cite{Cassel:2010px}. Recall 
that in MSSM $\lambda$ is  very small and  fixed by 
gauge interactions (at tree level) and this is  one  source for the above problems.}
  $v^2=-m^2/\lambda$ for $v\sim O(100 {\rm GeV})$, $m\sim O(1 {\rm TeV})$, that demands
a larger $\lambda$.
As a result one can
obtain a smaller  $\delta\chi^2\propto \ln\Delta_q$. As mentioned, the increased
complexity of the model (more parameters) is to be avoided, since then $n_{df}$ can decrease
and $\delta\chi^2/n_{df}$ may not change much (or even increase). 
The GNMSSM model is an example of i), where a supersymmetric mass term for the additional singlet 
essentially enabled a smaller  $\delta\chi^2/n_{df}$ than in other MSSM-like 
models\footnote{For a large value of the supersymmetric mass term of 
the singlet ($M$ of few TeV, 5-8), the correction to the higgs  effective quartic coupling $\lambda$ is
$\delta \lambda \sim (2\mu/M) \sin 2\beta$
already at tree level \cite{Cassel:2009ps}, with impact on $\Delta_q$ and $\delta\chi^2$.
\cite{Ross:2012nr}. For a recent study of the GNMSSM and
its LHC signatures see \cite{Dreiner:2012ec}.}.
Similar but milder effects exist in the NMSSM\footnote{
The correction to $\lambda$ and $m_h$ is in this 
case  restricted by perturbativity in the singlet coupling $\tilde\lambda$
(of $\tilde\lambda S H_1. H_2$) and also its proportionality
 to $\tilde\lambda\sin^2 2\beta$ instead, while in the GNMSSM is $\propto \sin2\beta$.}.
An increase of $\lambda$  could also  be generated by using idea ii)
by adding more gauge symmetry  (e.g. \cite{tait}). Regarding  option iii),
one can consider additional massive states that couple to the higgs sector, and
that in the low energy generate corrections to the higgs potential and 
effective $\lambda$ and $m_h$  \cite{MCDG} with similar effects.
 These ideas  may indicate  the direction for SUSY model building.

\medskip\noindent
{\bf b).} The above negative implications for some SUSY models remind us about the
real possibility that no sign of TeV-scale supersymmetry may be found at the LHC.
If so, this can suggest its scale  is significantly larger than few TeV.
 Alternatively,  one could attempt to forbid
the existence of {\it asymptotic} supersymmetric states  while trying to preserve
some of  the  nice advantages of SUSY.
In such scenario, superpartners would be present only as internal lines in 
loop diagrams, they  would not be real, asymptotic  final states. One 
could describe this situation  by using some variant of {\it non-linear} supersymmetry
that can be described in a superfield formalism endowed with constraints
(see  examples in \cite{Seiberg}).
The hope would be to  preserve  SUSY results like fixing the EW scale, 
gauge couplings unification, radiative EW symmetry breaking, for which the superpartners 
in the loops play a crucial role\footnote{A  related idea
exists \cite{manton}, based on  a similarity to gauge fixing 
in gauge theories and the subsequent
emergence of the ghost degrees of freedom of different statistics.
This would have  an analogue in the above SUSY scenario 
in ``fixing the gauge'' in the Grassmann space.
Similar to the emergence of ghosts in gauge theories as non-asymptotic states 
one could attempt to obtain non-asymptotic  superpartners states.}. 
However, it is difficult to  realize this idea in practice. One 
also recalls the supersymmetric quantum mechanics case where 
SUSY is used only as a  tool    for performing complex calculations
\cite{qm}, which could suggest ideas for the field theory case.

\medskip
\noindent
{\bf c).} The remaining possibility is to abandon (low-energy) SUSY 
 and eventually consider a different symmetry instead.
One can use model building based  on SM extended with 
 the (classical) scale  symmetry, thus forbidding a tree level higgs mass.
This symmetry is broken at the loop level by anomalous dimensions, which 
would  bring in only log-like dependence on the mass scales \cite{bardeen}. Additional  
requirements  (unitarity, etc) could be added. For model building along 
this direction see  some examples in \cite{Shaposhnikov:2008xi}  and references therein.

\section{Conclusions}

The main motivation for TeV-scale supersymmetry was to solve the hierarchy problem and therefore
fix the electroweak scale (vev $v$ or $m_Z$)  in the presence of the quantum corrections. 
Rather surprisingly, 
the  numerical methods  that evaluate the likelihood (or its $\chi^2\equiv-2\ln L$) to fit
 the data in SUSY models do not account for the  $\chi^2$ ``cost''  that is 
due to   fixing the EW scale to its measured value ($m_Z^0$). 
When this condition is properly  imposed, one finds that  $\chi^2$ receives 
a positive correction,  $\delta\chi^2 =2\ln\Delta_q > 0$, where $\Delta_q$ has some
resemblance to a ``traditional''  EW fine-tuning measure in 
``quadrature''\footnote{For this reason one can say that the ``traditional'' 
fine-tuning  is an intrinsic 
part of the likelihood to fit the data that includes the EW scale value ($m_Z^0$).}.
The correction $\delta\chi^2$  must  be included in the 
analysis of the total $\chi^2$ of the SUSY models.

Our analysis  also showed the contradiction that is present in those data fits of 
SUSY models that report a good fit $\chi^2/n_{df}\approx 1$ of the data {\it including}
the EW scale itself  while at the same time have a large EW fine-tuning.
A large fine-tuning  suggests
a significant variation of the EW scale ($m_Z$) away from the measured value ($m_Z^0$)
under a small variation of the SUSY parameters;
this impacts on the value of total $\chi^2/n_{df}$ and  worsens it, 
in contradiction with its good value (total $\chi^2/n_{df}\approx 1$)
that is often reported in such data fits.
The solution to this puzzle was  mentioned above:
our claim is that in these data fits the
likelihood to fit the data ($\chi^2$)  does not account for what we identified as the 
 $\delta\chi^2$ ``cost'' (with $\delta\chi^2/n_{df}>1$)
 of fixing the EW scale to its measured value. 
For this reason current data fits {\it underestimate}  the total value 
of $\chi^2/n_{df}$  in SUSY models.

For the recently measured value of the higgs mass ($\approx 126$ GeV),
 the correction $\delta\chi^2$ was estimated and
was shown to be  significant in most popular SUSY models: constrained MSSM (CMSSM),
models with non-universal  higgs soft masses (NUHM1, NUHM2) or
with non-universal gaugino masses (NUGM) and in the NMSSM and it was milder in the
 generalized version of NMSSM (GNMSSM).
This correction has negative implications for the data fits of SUSY models.
Our estimates show that for $m_h\approx 126$ GeV, 
this correction alone is $\delta\chi^2/n_{df}> 1.5$, 
which violates the traditional condition for a good fit 
already  before fitting observables other than the EW scale. 
Adding  this  contribution to that due to these observables
assumed to bring the ``usual''  $\chi^2/n_{df}\approx 1$,  would give a {\it total}
 $\chi^2_w/n_{df}=(\chi^2+\delta\chi^2)/n_{df}>2.5$,  hardly acceptable.
Another way to express this result is  that in these models 
a good fit of $m_Z^0$ and current EW data (i.e. $\chi^2_w/n_{df}\approx 1$),
and a simultaneous large EW fine tuning (i.e. $\delta\chi^2/n_{df}=2\ln\Delta_q/n_{df}> 1$) 
are not simultaneously possible.
Further contributions  to $\delta\chi^2$ also exist from the
 Yukawa couplings, but their effect was not discussed in this work.
Let us mention that
these results rely on the assumption made in the calculation of 
the total $\chi^2_w$ that  the EW scale ($m_Z$) 
is independent of the other observables.  Therefore it is possible  
that correlations effects  between  these  can modify 
the above result  and  even relax the upper value of  $\Delta_q$
that is still consistent with  a good fit ($\chi^2_w/n_{df}\approx 1$).

To conclude, the requirement of fixing the EW scale in SUSY models
 brings a correction to the likelihood ($\chi^2$) that can have significant
 negative implications for  the quality  of the data fits.
We provided an estimate of this correction  and argued that it 
must be included in  the total likelihood to fit the EW data, 
when  testing the  viability of  SUSY models.

\bigskip\bigskip\bigskip\bigskip
\noindent{\bf Acknowledgements:\,\,}
The author thanks Hyun Min Lee (KIAS Seoul, Korea) for discussions on this topic.
 This work  was supported by a grant of the Romanian National Authority for Scientific
Research, CNCS - UEFISCDI, project number PN-II-ID-PCE-2011-3-0607.


\end{document}